\documentclass[aps,amsfonts,preprint]{revtex4}
\def\S2{\bar{S}}
\def\a{\alpha}
\def\b{\beta}

\def\P{\Pi}

\def\o{\omega}
\def\t{\tau}
\def\an{a_{n}}
\def\ta{\tilde{a}_n}
\def\and{a_{n}^\dagger}

\def\tab{\tilde{\bar{a}}_n}
\def\on{\omega_{n}}
\def\q{S_{n}}
\def\tq{\tilde{S}_n}

\def\x{\S2_{n}}
\def\tx{\tilde{\S2}_n}
\def\sn2d{\Sn2^\dagger}
\def\({\left(}
\def\){\right)}
\def\<{\left\langle}
\def\>{\right\rangle}
\def\s{\sigma}

\begin{document}
\title{ Perspectives of TFD on String Theory }

\author{M. C. B. Abdalla{\footnote{daniel@ift.unesp.br}}, 
 A. L. Gadelha {\footnote{gadelha@ift.unesp}}
and Daniel L. Nedel{\footnote{daniel@ift.unesp}}}

\affiliation{Instituto de F\'{\i}sica Te\'{o}rica, Unesp, Pamplona 145,
S\~{a}o Paulo, SP, 01405-900, Brazil }

\begin{abstract}
Considering the type IIB superstring in a pp wave background
some recent ideas and perspectives of Thermo Field Dynamics on string
theory are presented. The thermal Fock space is constructed attempting to
consider a possible finite temperature version of the BMN correspondence
in this framework. Also, the thermal vacuum is identified as a string
boundary state realizing the thermal torus interpretation in the ambit
of Thermo Field Dynamics. Such a interpretation consists of a
generalization of some recent analysis for the closed bosonic string.

\end{abstract}

\maketitle




For the last 20 years, string theory has been considered the best
candidate for a theory that quantizes gravity. The perturbative
sector of the theory was explored in order to place the Yang-Mills
and gravitation theory at the same footing. The most impressive
characteristic of string theory comes from the fact that the
structure of the pertubative string theory is much more
constrained than the one of field theory. This constrained
structure gives the dimension of the space time, the gauge group
and demands supersymmetry. Also, the formulation of the
perturbative string at finite temperature is constrained. There is
a temperature at above which the statistical partition function
diverges (Hagedorn temperature) \cite{Witten2}. The very
existence of the Hagedorn temperature shows that the fundamental
degrees of freedom of the string theory could not be the ones of
the perturbative string. In this direction,  the perturbative
string theory at finite temperature  may provides important pieces
of evidence of the true degrees of freedom of the theory at
non-perturbative regime (higher temperatures). However, an
understanding of the Hagedorn temperature, and the implications of
this constrain in the space-time physics is still lacking. This
lack of understanding motivates the study of new methods to
introduce temperature in string theory. For example, it is known
that the Hagedorn temperature comes from the exponential growth of
states as function of energy. In this case the passage of the
microcanonical ensemble to gran-canonical ensemble of statistical
mechanics is absolutely non-trivial and it is hard to see where the
results of one formalism is valid in the other.
In this sense a more general formalism to deal with
systems at finite temperature could be very useful. It is the case
of the formalism that is going to be
presented here, named Thermo Field Dynamics (TFD). The TFD was
developed  by Takahashi and Umezawa in order to handle finite
temperature with a real time operator formalism \cite{UME}.
The main idea is to interpret the statistical average of an
operator $Q$, $\left\langle  Q \right\rangle$  as the expectation
value in a thermal vacuum
\begin{equation}
\left\langle  Q \right\rangle  = \left\langle 0 (\b ) \left| Q
\right| 0 (\b )\right\rangle,
\label{tr1}
\end{equation}
for $\beta=1/T$ where $T$ is the temperature.

The thermal vacuum is constructed by means of a Bogoliubov
transformation. Also, the creation and annihilation operators are
transformed in order to construct a thermal Fock space. All the
thermodynamics quantities can be defined as matrix elements of an
operator in the thermal vacuum. The thermal effects are consequence
of the condensed state structure of that thermal vacuum.

Concerning string theory, the idea of building a thermal Fock
space can be very fruitful. For example, the ``coherent state''
description of $D_{p}$-branes in the perturbative limit of the
theory, where they are viewed as a boundary state constructed in
the Fock space of the closed string \cite{divec}, made such kind
of approach particularly tempting. In fact, in this context, the
TFD has been used in order to study thermal properties of bosonic
$D_{p}$-branes \cite{IVV,AGV2,AEG,AGV3,AGV4,AGV5}. Another
important example rests in the ADS/CFT correspondence. More
specifically when one wants to study thermodynamics of type IIB
superstring propagating in a pp wave background. In this case, the
operatorial characteristic of TFD is suitable \cite{nos}, since
the so called Berenstein-Maldacena-Nastase (BMN) correspondence
\cite{BMN} can be explored at finite temperature directly in the
Fock space.

%
%
In this work, some recent ideas and perspectives of the TFD
approach on string theory are presented concentrating on the study
of type IIB superstring in a pp-wave background at finite
temperature. This system is worked on in order to present some
perspective on the understanding of a thermal BMN correspondence
and at the end an
interpretation for the thermal vacuum is made in terms of a
particular boundary state for the system under consideration.
This interpretation allows one to understand the relationship between
imaginary time \cite{adas} and TFD when both formalisms are used to
study type IIB superstring  at finite temperature. It consists of
a generalization, to include fermionic degrees of freedom in a
recent analysis for the closed bosonic string \cite{torus}.

In the last years the BMN conjecture \cite{BMN}, was explored to
give some insights into the strings thermodynamics in terms of Yang
Mills thermodynamics \cite{Greene,Br}. In this conjecture
the string propagates in a pp wave background. Such a background
is obtained as the Penrose limit of $ADS_{5}\times S^5$
\cite{Blau2}, where the only surviving components of the
Ramond-Ramond five form are: $F_{+1234}= F_{+5678}=\mu$. On the
gauge theory side the limit focuses on a set of operators which
have R charges $J$ and the conformal dimension $\Delta$ satisfying
$J\approx\sqrt{N}$ and $\Delta\approx J$, for fixed Yang-Mills
coupling and $N$ going to infinity. This set up provides a complete
dictionary between self-states of the string hamiltonian and
Yang-Mills operators with $\Delta -J$ charge. For example, the
vacuum in the string side has zero energy and it is related to an
operator in the gauge side with zero value for $\Delta - J$
\begin{equation}
\left|0,p^+\> \rightarrow O^J(0) \left|vac\>, \qquad
O(x)=\frac{1}{\sqrt{JN^J}}TrZ^J, \label{dic}
\end{equation}
where $\left|vac\>$ is the Yang-Mills vacuum and $O^J$ is composed
of two out of the six scalar fields of the ${\cal N}=4$ super Yang
Mills multiplet: $ Z= \frac{1}{2}\(\phi^5 + i\phi^6\)$. The trace
is taken over the $SU(N)$ index. Lately there have been some
interesting works studying finite temperature effects of type IIB
superstring in the pp waves background \cite{Zayas,Suga,Grig}. In
a general way these works used the imaginary time formalism, to
compute the superstring partition function and the free energy on
a torus. However, using this formalism is difficult to take into
account in the statistical average, only sectors that survive to
the Penrose limit. As a consequence it is hard to see if the BMN
correspondence works at finite temperature. On the order hand,
with the TFD approach it is possible to construct a thermal Fock
space for both sides of the conjecture and verify directly in this
Fock space if the dictionary $\(\ref{dic}\)$ works.

The solutions of the equations of motion with periodic boundary
conditions for the type II superstring in pp wave background
are \cite{Met2}
\begin{eqnarray}
X^I = x^I_0 \cos(m\t)+\frac{\a^\prime}{m}p^I_0 \sin(m\t)\
+\sqrt{\frac{\a^\prime}{2}} \sum_{n > 0}\frac{1}{\sqrt{\o_n}}
\left[\(\an^I e^{-i(\on\t-k_n\s)}+\an^{\dagger \: I} e^{i(\on\t -
k_n\s)}\) \right. \nonumber
\\
\left. +\({\bar a}_{n}^{I} e^{-i(\on\t+k_n\s)}+{\bar
a}_{n}^{\dagger \: I} e^{i(\on\t + k_n\s)}\)\right],
\end{eqnarray}
and
\begin{eqnarray}
S^a = \cos(m\t)S_0^a + \sin(m\t)\P_{ab} \S2_0^b + \sum_{n > 0}
c_{n} \left[\q^{a} e^{-i(\on\t-k_n\s)}\right. \left. +
S_{n}^{\dagger a} e^{i\( \o_{n}\t-k_{n}\s\)}\right. \nonumber
\\
+ \left. i\frac{\on-k_n}{m}\P_{ab}\(\S2_{n}^{b}
e^{-i(\on\t+k_n\s)}-\S2_{n}^{\dagger b} e^{i\(\o_{n}\t+k_{n}\s\)}
\)\right],
\end{eqnarray}
\begin{eqnarray}
\S2^a = \cos(m\t)\S2_{0}^a - \sin(m\t)\P_{ab} S_{0}^b + \sum_{n >
0}c_{n}\left[\S2_{n}^{a} e^{-i(\on\t+k_n\s)}\right. \left. +
\S2_{n}^{\dagger a} e^{i\( \o_{n}\t+k_{n}\s\)}\right. \nonumber
\\
- \left. i\frac{\on-k_n}{m}\P_{ab}\(S_{n}^{b} e^{-i(\on\t-k_n\s)}-
S_{n}^{\dagger b} e^{i\(\o_{n}\t-k_{n}\s\)}
 \)\right],
\end{eqnarray}
where we set $m=\mu\ \a^{\prime} p^{+}$ and
\begin{equation}
\o_n = \sqrt{m^2 + k_n^2},\qquad c_n
=\frac{1}{\sqrt{1+(\frac{\o_{n} - k_{n}}{m})^2}},\qquad k_n =2\pi
n.
\end{equation}
The canonical quantization gives the standard commutator and
anti-commutator relations of harmonic oscillator and the zero mode
part is written as follows
\begin{eqnarray}
a_0^I &=& \frac{1}{\sqrt{2m}}(p_0^I -imx_0^I),\qquad
a_0^{\dagger\:I}=
\frac{1}{\sqrt{2m}}(p_0^I +imx_0^I),\nonumber \\
S_{\pm}^a &=&\frac{1}{2}\(1\pm \P\)_{ab}\frac{1}{\sqrt{2m}}\(S_0^b
\pm i\S2_0^b\),\qquad S_{\pm}^{\dagger\:a}= \frac{1}{2}\(1\pm
\P\)_{ab}\frac{1}{\sqrt{2m}}\(S_0^b \mp i\S2_0^b\).
\end{eqnarray}
The vacuum $\left|0,p^{+}\>$ is defined by
\begin{eqnarray}
\q\left|0,p^+\>&=&\x\left|0,p^+\>=0, \qquad n>0,
\nonumber \\
a_n^I\left|0,p^+\>&=&{\bar a}_n^I\left|0,p^+\>=0, \qquad n>0,
\nonumber \\
S_{\pm}\left|0,p^+\>&=&a_{0}^{I}\left|0,p^+\>=0.
\end{eqnarray}

Let us to introduce the temperature in this system using TFD. The
TFD algorithm starts by duplicating the degrees of freedom. To this
end a  copy of the original Hilbert space is introduced and denoted
by $\widetilde{H}$. The tilde Hilbert space is built with a set of
oscillators: $\tilde{a}_0$, $\tilde{S}_{\pm}$, $\ta$, $\tab$,
$\tq$, $\tx$ that have the same (anti-) commutation properties as
the original ones. The operators of the two systems (anti-) commute
among themselves and the total Hilbert space is the tensor product
of the two spaces. The map between the tilde and non-tilde operators
is defined by the following tilde (or dual) conjugation rules
\cite{UME,kha5}
\begin{eqnarray}
(A_{i}A_{j})\widetilde{} &=&\widetilde{A}_{i}\widetilde{A}_{j},
\qquad  
(A_{i}^{\dagger })\widetilde{},
=\sigma(\widetilde{A}_{i})^{\dagger }, 
\qquad  (\widetilde{A}_{i})\widetilde{}
=A_{i},
\nonumber \\
(cA_{i}+A_{j})\widetilde{} &=& c^{\ast
}\sigma\widetilde{A}_{i}+\sigma\widetilde{A}_{j}, \qquad
[\widetilde{A}_{i},A_{j}] =0,
\label{til}
\end{eqnarray}
where $\sigma = 1$ for bosons, $\sigma=-1$ for fermions
and $c \in \mathbb{C}$.
From these rules, the tilde system can be describe
by a string that propagates backwards in imaginary time \cite{torus}.
So, the TFD starts defining two independent free strings, defining
two cylinders.

We can now construct the thermal vacuum. This is achieved by
implementing a Bogoliubov transformation in the total Hilbert
space. The transformation generator is given by
\begin{equation}
G=G^{B}+G^{F}, \label{gen}
\end{equation}
\begin{eqnarray}
G^{B}&=&G_{0}^{B} + \sum_{n=1} \(G_{n}^{B} + {\bar G}_n^{B}\),
\label{genb}
\\
G^{F}&=&G_{+}^{F} + G_{-}^{F} + \sum_{n=1} \(G_{n}^{F} + {\bar
G}_{n}^{F}\), \label{genf}
\end{eqnarray}
where
\begin{eqnarray}
G_{0}^{B}&=&-i\theta_{0}^{B}\(a_{0}\cdot {\tilde a}_{0}-{\tilde
a}_{0}^{\dagger} \cdot a_{0}^{\dagger}\), \label{gen0b}
\\
G_{n}^{B}&=&-i\theta_{n}^{B}\(a_{n}\cdot {\tilde a}_{n}-{\tilde
a}_{n}^{\dagger} \cdot a_{n}^{\dagger}\), \label{genbn}
\\
{\bar G}_{n}^{B}&=&-i{\bar \theta}_{n}^{B}\({\bar a}_{n}\cdot
{\tilde {\bar a}}_{n}-{\tilde {\bar a}}_{n}^{\dagger} \cdot {\bar
a}_{n}^{\dagger}\), \label{genbnb}
\\
G_{\pm}^{F}&=&-i\theta_{\pm}^{F}\({\widetilde S}_{\pm}\cdot
S_{\pm}-S_{\pm}^{\dagger} \cdot {\widetilde S}_{\pm}^{\dagger}\),
\label{gen0f}
\\
G_{n}^{F}&=&-i\theta_{n}^{F}\({\widetilde S}_{n}\cdot
S_{n}-S_{n}^{\dagger} \cdot {\widetilde S}_{n}^{\dagger}\),
\label{genfn}
\\
G_{n}^{F}&=&-i{\bar \theta}_{n}^{F}\({\widetilde {\bar
S}}_{n}\cdot {\bar S}_{n}-{\bar S}_{n}^{\dagger} \cdot {\widetilde
{\bar S}}_{n}^{\dagger}\). \label{genfnb}
\end{eqnarray}
Here, the $B$ and $F$ labels specify fermions and bosons, the
dots represent the inner products and $\theta$, ${\bar \theta}$
are parameters that depend on temperature. The thermal vacuum is
defined by the following relation
\begin{equation}
\left |0\(\theta\)\right\rangle = e^{-i{G}}\left.
\left|0\right\rangle \!\right\rangle \nonumber
\label{vac}
\end{equation}

The effect of the Bogoliubov transformation is to entangle the
elements of the two Hilbert spaces. After the transformation, the
image of the two independent strings  is lost. The creation and
annihilation operators at $T \neq 0$ are given by the Bogoliubov
transformation as follows
\begin{eqnarray}
a_{n}^{I}\(\theta_{n}\)&=&e^{-iG}a_{n}^{I}e^{iG}
=\cosh\(\theta_{n}\)a_{n}^{I} - \sinh\(\theta\){\widetilde
a}_{n}^{\dagger \: I},
\\
{S}^{a}_{n}\(\theta_{n}\)&=&e^{-iG}{S}^{a}_{n}e^{iG}
=\cos\(\theta_{n}\){S^ a}_{n}
- \sin\(\theta_{n}\){\widetilde {S}}^{\dagger\:a}_{n},
\end{eqnarray}
and the same for the other operators. These operators annihilate
the state written in $\(\ref{vac}\)$ defining it as the vacuum.
The thermal Fock space is constructed by applying the thermal
creation operators to the vacuum $\(\ref{vac}\)$. As the
Bogoliubov transformation is canonical, the thermal operators obey
the same commutation relations as the operators at $T=0$.

Next, we show how the thermodynamical quantities can be derived
introducing first a free energy like potential defined by
\begin{equation}
\mathcal{F}=\mathcal{E}-\frac{1}{\beta }\mathcal{S},
\end{equation}
where $\mathcal{E}$ is related with the thermal
energy and $\mathcal{S}$ with the entropy of the string. In TFD the
thermal energy is given by computing the matrix elements of the $T=0$
hamiltonian in the thermal vacuum. In order to take into account
the level matching condition, the shifted hamiltonian is used
\begin{equation}
\mathcal{E}=
\left\langle 0(\theta )\left| H_s \right|0(\theta)\right\rangle
=\left\langle 0 (\theta ) \left| H
+ \frac{i\lambda}{\beta}P \right| 0 (\theta)\right\rangle.
\end{equation}
Here $\lambda$ is a lagrange multiplier that fixes the $S^1$
isometry of the closed string and $P$ is the worldsheet momentum.
The entropy of the superstring is calculated by evaluating the
expected value of the entropy operator defined in \cite{nos} in
the thermal vacuum. By minimizing the potential $\mathcal{F}$ with
respect to $\theta$ we find the explicit dependence of these
parameters in relation to $\omega_{n}$, $\beta$ and $\lambda$.
In this way we have
\begin{equation}
\sinh^{2}\(\theta_{0}^{B}\)=\frac{1}{e^{\frac{\beta m}{p+}}-1},
\qquad \sin^{2}\({\theta}_{\pm}^{F}\)=\frac{1}{e^{\frac{\beta
m}{p+}}-1},
\end{equation}
for the zero modes, and
\begin{eqnarray}
\sinh^{2}\(\theta_{n}^{B}\)&=&\frac{1}{e^{\frac{\beta
\omega_{n}}{p+}+ i\lambda k_{n}}-1}, \qquad \sinh^{2}\(
\overline{\theta}_{n}^{B}\)=\frac{1}{e^{\frac{\beta\omega_{n}}{p+}-
i\lambda k_{n}}-1},
\nonumber
\\
\sin^{2}\(\theta_{n}^{F}\)&=&\frac{1}{e^{\frac{\beta
\omega_{n}}{p+}+ i\lambda k_{n}}+1}, \qquad \sin^{2}\(
\overline{\theta}_{n}^{F}\)=\frac{1}{e^{\frac{\beta\omega_{n}}{p+}-
i\lambda k_{n}}+1},
\label{dist}
\end{eqnarray}
for the other modes. With these results, one can write the free energy
like potential as \cite{nos}
\begin{equation}
\mathcal{F}\(\lambda,\b/p^{+}\)=-\frac{1}{\beta} \ln \prod_{n={\mathbb Z}}
\left[ \frac{1+e^{-\frac {\beta \omega_{n}}{p+}+ i \lambda k_{n}}}
{1-e^{-\frac {\beta \omega_{n}}{p+}+i \lambda k_{n}}}\right]^{8}=
-\frac{1}{\beta}\ln \(z_{lc}(\b / p^+,\lambda)\), \label{final}
\end{equation}
where $z_{lc}(\b / p^+,\lambda)$ is the torus transverse partition
function calculated in \cite{Zayas}.

So, the thermal vacuum reproduces the same results of the imaginary
time formalism, when the worldsheet is defined on a torus with
moduli space parameters $\t=\lambda+i\t\frac{\b}{2\pi}$. The
next step would be to construct a thermal vacuum for the Yang
Mills side and  verify the BNM dictionary directly in the thermal
Fock space. However, before going on with this program a question
needs to be answered; How did the torus defined by the moduli space
parameter $\t=\lambda+i\t\frac{\b}{2\pi}$ appears in the TFD approach?

The TFD approach starts with a tree level string and an auxiliary
string (tilde system). Suppose we want to construct a torus with
the two cylinders defined by the strings. As the partition
function for the torus is defined after Wick rotation, we need to
go to Euclidean time $\(\tau=-it\)$. In addiction, the tilde string
is a mirror of the string and propagates backwards in imaginary time.
So, when the string propagates from zero to a euclidian time $\b/2$
for example, the tilde string propagates from zero to $-\b/2$ in
euclidian time. Then, a torus can be constructed by gluing
together the end of the original cylinder with the origin of the
tilde one, and vice-versa \cite{torus}. Also, before gluing, the
identification $\tilde{\sigma}=\sigma-\pi\lambda$, must be done in
order to take into account the Dehn twist in one cycle. The two
parameters of the resulting torus moduli space will be related to
$\b $ and $\lambda$. The above considerations can be written as
follows
\begin{eqnarray}
X\(t,\sigma\)
-\widetilde{X}\(-t-\frac{\b}{2},\sigma-\lambda\pi\)&=&0,
\qquad
X^I\(-\tilde{t}-\frac{\beta}{2},\tilde{\sigma}+\lambda\pi\)
-\widetilde{X}^I\(\tilde{t},\tilde{\sigma}\)=0,
\nonumber \\
S^a\(t,\sigma\)
-\widetilde{S}^a\(-t-\frac{\b}{2},\sigma-\lambda\pi\)&=&0,
\qquad
S^a\(-\tilde{t}-\frac{\beta}{2},\tilde{\sigma}+\lambda\pi\)
-\widetilde{S}^a\(\tilde{t},\tilde{\sigma}\)=0. \label{ident}
\end{eqnarray}
Expanding $X\(t,\sigma\)$, $S\(t,\sigma\)$ and
$\widetilde{X}\(\widetilde {t}, \widetilde{\sigma}\)$,
$\widetilde{S}\(t,\sigma\)$ in modes, the above identification
turns out to be a set of operatorial equations for a boundary
state $|\Phi \rangle =|\phi,\widetilde{\phi }\rangle $. Resolving
the boundary state equations we get the following normalized
solution
\begin{eqnarray}
|\Phi \rangle &=& \left(
\frac{1}{\cosh(\theta_{0}^{B})}\right)^{8}
\left(\cos(\theta_{+}^{F})\right)^{4}\left(\cos(\theta_{-}^{F})\right)^{4}
e^{\tanh\(\theta_{0}^{B}\)\(a_{0}^{\dagger}\cdot {\tilde
a}_{0}^{\dagger}\)}e^{\tan\(\theta_{+}^{F}\)\(S_{+}^{\dagger}
\cdot {\widetilde S}_{+}^{\dagger}\) +
\tan\(\theta_{+}^{F}\)\(S_{-}^{\dagger} \cdot {\widetilde
S}_{-}^{\dagger}\)} \nonumber
\\
& \times &\prod_{n=1}\left[\left(
\frac{1}{\cosh(\theta_{n}^{B})}\right)^{8}\left(
\frac{1}{\cosh({\bar \theta}_{n}^{B})}\right)^{8}
e^{\tanh\(\theta_{n}^{B}\)\(a_{n}^{\dagger}\cdot {\tilde
a}_{n}^{\dagger}\)+ \tanh\({\bar \theta}_{n}^{B}\)\({\bar
a}_{n}^{\dagger}\cdot {\tilde {\bar a}}_{n}^{\dagger}\)}\right.
\nonumber
\\
&\times&
\left.\left(\cos(\theta_{n}^{F})\right)^{8}\left(\cos({\bar
\theta}_{n}^{F})\right)^{8}e^{\tan\(\theta_{n}^{F}\)\(S_{n}^{\dagger}
\cdot {\widetilde S}_{n}^{\dagger}\) + \tan \({\bar
\theta}_{n}^{F}\)\({\bar S}_{n}^{\dagger} \cdot {\widetilde {\bar
S}}_{n}^{\dagger}\)}\right] \left.
\left|0\right\rangle\!\right\rangle \label{tva}.
\end{eqnarray}
This is precisely the thermal state $\(\ref{vac}\)$ when the
relations $\(\ref{dist}\)$ are used and the Bogoliubov generators
are expanded.

Finally, we can conclude that the effect of the Bogoliubov
transformation is to entangle the two strings in order to make a
torus in such away that the worldsheet fields are confined in a
restricted region $\beta$ of the time axis. Such a confinement
was pointed out in the scope of field theory and Casimir effect
\cite{adecas}. The thermal state is a boundary state responsible
for this confinement. This is a very suggestive result, since it
opens the possibility of other kinds of boundary states
(D-branes for example) to appear at some higher temperature.
It will be interesting to relate this topological interpretation
for the thermal vacuum constructed for this type II superstring,
to a thermal boundary state in the Yang Mills side of the BMN
conjecture. This work is in progress. In other direction, we
can use the operatorial method developed in \cite{Gaume} to study
string at finite temperature on higher genus surfaces.
\section*{Acknowledgements}
We would like to thank D. Z. Marchioro and I. V. Vancea for useful
discussions. M. C. B. A. was partially supported by the CNPq Grant
302019/2003-0, A. L. G. and D. L. N. are supported by a FAPESP
post-doc fellowship.


\end{document}